\documentclass[
aps,
pra,
twocolumn,
superscriptaddress,
floatfix
longbibliography
]{revtex4-1}
\usepackage[dvipdfmx]{graphicx}
\usepackage{verbatim}
\usepackage{epsf}
\usepackage{natbib}
\usepackage{dcolumn}
\usepackage{bm}
\usepackage{color} 
\usepackage{ulem}
\usepackage{amsmath}
\usepackage{amssymb}
\usepackage{txfonts}
\usepackage[
colorlinks=true,
citecolor=blue,
urlcolor=blue,
setpagesize=false]{hyperref}
\newcommand{\bs}   {\boldsymbol}

\newcommand{\mcal} {\mathcal}
\newcommand{\imag} {{i}}
\newcommand{\dd}   {{d}}
\newcommand{\e}    {{e}}

\newcommand{\up}   {\uparrow}
\newcommand{\dn}   {\downarrow}

\newcommand{\s}    {\sigma}
\newcommand{\w}    {\omega}

\newcommand{\eps}  {\epsilon}

\newcommand{\ksum} {\frac{1}{N_{\rm cell}} \sum_{\bs{k}}}
\newcommand{\ksumhalf} {\frac{1}{2N_{\rm cell}} \sum_{\bs{k}}}
\newcommand{\ksumZhalf} {\frac{Z}{2N_{\rm cell}} \sum_{\bs{k}}}
\newcommand{\kint} {\frac{S_{\rm cell}}{(2\pi)^2} \int \dd^2 \bs{k}}

\newcommand{\mytitle}{
  Fermi-liquid ground state of interacting Dirac fermions in two dimensions 
}
  
\hypersetup{
  colorlinks=true,
  linkcolor=[rgb]{0.60,0.00,0.00},
  citecolor=[rgb]{0.0,0.0,0.60},
  urlcolor=[rgb]{0.0,0.0,0.60},
  setpagesize=false
}

\begin{document}

\title{\mytitle}

\author{Kazuhiro~Seki}
\affiliation{SISSA--International School for Advanced Studies, Via Bonomea 265, 34136, Trieste, Italy} 
\affiliation{Computational Materials Science Research Team, RIKEN Center for Computational Science (R-CCS), Kobe, Hyogo 650-0047,  Japan}
\affiliation{Computational Condensed Matter Physics Laboratory, RIKEN Cluster for Pioneering Research (CPR), Wako, Saitama 351-0198, Japan}

\author{Yuichi~Otsuka}
\affiliation{Computational Materials Science Research Team, RIKEN Center for Computational Science (R-CCS), Kobe, Hyogo 650-0047,  Japan}

\author{Seiji~Yunoki}
\affiliation{Computational Materials Science Research Team, RIKEN Center for Computational Science (R-CCS), Kobe, Hyogo 650-0047,  Japan}
\affiliation{Computational Condensed Matter Physics Laboratory, RIKEN Cluster for Pioneering Research (CPR), Wako, Saitama 351-0198, Japan}
\affiliation{Computational Quantum Matter Research Team, RIKEN Center for Emergent Matter Science (CEMS), Wako, Saitama 351-0198, Japan}
            
\author{Sandro~Sorella}
\affiliation{SISSA--International School for Advanced Studies, Via Bonomea 265, 34136, Trieste, Italy} 
\affiliation{Computational Materials Science Research Team, RIKEN Center for Computational Science (R-CCS), Kobe, Hyogo 650-0047,  Japan}

\begin{abstract}
  An unbiased zero-temperature auxiliary-field quantum Monte Carlo method is employed to 
  analyze the nature of the semimetallic phase of the 
  two-dimensional Hubbard model on the honeycomb lattice at half filling. 
  It is shown that the quasiparticle weight $Z$ 
  of the massless Dirac fermions at the Fermi level, 
  which characterizes the coherence of zero-energy single-particle excitations, 
  can be evaluated in terms of the long-distance equal-time single-particle Green's function. 
  If this quantity remains finite in the thermodynamic limit, 
  the low-energy single-particle excitations of the correlated semimetallic phase  
  are described by a Fermi-liquid-type single-particle Green's function. 
  Based on the unprecedentedly large-scale numerical simulations on finite-size clusters containing 
  more than ten thousands 
  sites, we show that the quasiparticle weight 
  remains finite in the semimetallic phase 
  below a critical interaction strength. 
  This is also supported by the long-distance algebraic behavior ($\sim r^{-2}$, where $r$ is distance) 
  of the equal-time single-particle Green's function that is expected for the Fermi liquid. 
  Our result thus provides 
  a numerical confirmation of 
  Fermi-liquid theory in two-dimensional correlated metals. 
\end{abstract}

\date{\today}

\maketitle

\section{Introduction}
The characterization of different  phases of matter is 
one of the essential issues in solid state physics. 
In the field of strongly correlated electrons, 
the correlation-induced metal-insulator transition~\cite{Imada1998} 
is of particular importance since the 
itinerancy and localization of electrons~\cite{Kohn1964,Resta1999} 
can be regarded 
as a many-electron realization of the wave-particle duality, 
the fundamental concept of quantum mechanics.

The Hubbard model~\cite{Gutzwiller1963,Kanamori1963,Hubbard1963} is 
certainly one of the most important models in condensed matter physics 
since it has inspired many ideas and led to milestone achievements 
for understanding the fascinating properties of the 
metal-insulator transition.  
In particular, a 
semimetal-insulator transition 
occurs in the Hubbard model in a certain class of lattices 
where massless Dirac-like dispersion appears in the noninteracting limit,
and has been therefore a subject 
of intense activity in recent years. 
Since such models can be constructed on bipartite lattices and thereby they are 
free from the negative-sign problem,
the numerically exact auxiliary-field quantum Monte Carlo (AFQMC) method 
has played a major role in the study of 
this semimetal-insulator transition.
 In order to determine the ground-state phase diagram, most of the previous calculations have focused on 
the order parameters in the insulating phase, including 
the single-particle excitation gap and the antiferromagnetic spin-structure factor~\cite{Meng2010,Sorella2012,Chang2012,Assaad2013}. 
Variants of such models have been further extended recently 
by coupling interacting Dirac fermions to Ising spins~\cite{Sato2017} or 
by introducing disordered transfer integrals~\cite{Ma2018}.

On the theoretical side, 
the Green's-function-based formalism~\cite{Migdal1957,Luttinger1961,Nozieres1962,Luttinger1962,AGD} 
of the Fermi-liquid theory~\cite{Landau1956} 
argues
that one of the most important characteristics in a correlated metallic state 
is the quasiparticle weight $Z$ at the Fermi level, 
because finite $Z$ implies the existence of coherent 
zero-energy single-particle excitations. 
Although massless Dirac fermions exhibit only Fermi points instead of full Fermi surfaces, 
the quasiparticle weight $Z$ remains well defined~\cite{Herbut2009}, 
despite that the low-energy single-particle excitations and the electronic transport 
can be substantially different from those in simple metals~\cite{CastroNeto2009,DasSarma2010}.  
In principle, $Z$ can be estimated from the imaginary-time-displaced single-particle Green's function 
at the Dirac point with the AFQMC method~\cite{Feldbacher2001,Assaad_book}. 
However, the computation of 
imaginary-time-displaced quantities is considerably more expensive and 
suffers from much larger signal-to-noise ratio 
than the corresponding equal-time correlations. 
This is probably  the main reason for preventing 
the calculation of $Z$ 
in the semimetallic phase with the AFQMC technique.
In this regard, recently, three of us~\cite{Otsuka2016} 
elucidated the quantum criticality 
emerging from  the continuous semimetal-insulator transition, 
with large-scale zero-temperature 
AFQMC simulations~\cite{Sorella1988,Sorella1989,Sorella1992,Sorella2012,Becca_Sorella_book}.
However, no direct and systematic calculation of the quasiparticle weight 
for interacting Dirac fermions has been reported yet. 
It should also be noted that, in spite of the recent development of various numerical techniques 
and the continuous improvement of computer performances,
a solid numerical evidence of the presence of quasiparticles and, by consequence,
a clear validation of the Fermi-liquid theory,
are still lacking for interacting fermions on any two-dimensional lattices. 

In this paper, we first show that the quasiparticle weight $Z$ 
of the massless Dirac fermions at the Fermi level 
can be evaluated from the ratio of the interacting and noninteracting 
equal-time single-particle Green's functions in the long-distance limit.  
The scheme is then demonstrated with the unbiased zero-temperature 
AFQMC simulation for the Hubbard model on unprecedentedly large finite-size clusters 
of the honeycomb lattice at half filling. 
Based on the numerical results for the quasiparticle weight, 
we address a fundamental and long-standing issue:  
whether the Fermi liquid can be realized in two spatial dimensions~\cite{Varma1989,Anderson1990,Anderson1991}. 
Our result implies that 
the Fermi-liquid picture is valid 
in the correlated semimetallic phase.

The rest of the paper is organized as follows. 
In Sec.~\ref{sec2}, we define the Hubbard model on the honeycomb lattice 
and describe the AFQMC method. 
In Sec.~\ref{sec3}, based on the Fermi-liquid theory, 
we show that the quasiparticle weight $Z$ of interacting massless Dirac fermions  
is calculated from the equal-time single-particle Green's function. 
In Sec.~\ref{sec4}, 
we provide the numerical results 
which strongly support the Fermi-liquid behavior in the semimetallic phase. 
In Sec.~\ref{sec5}, we summarize the paper and discuss 
the non-Fermi-liquid behavior in graphene. 
In Appendixes~\ref{appA} and \ref{appB}, 
we analyze 
the long-distance behavior of the equal-time Green's function 
in the semimetallic and insulating phases, 
respectively.

\section{Model and Method}\label{sec2}
\subsection{Hubbard model on the honeycomb lattice}

\begin{figure}
  \begin{center}
    \includegraphics[width=.85\columnwidth]{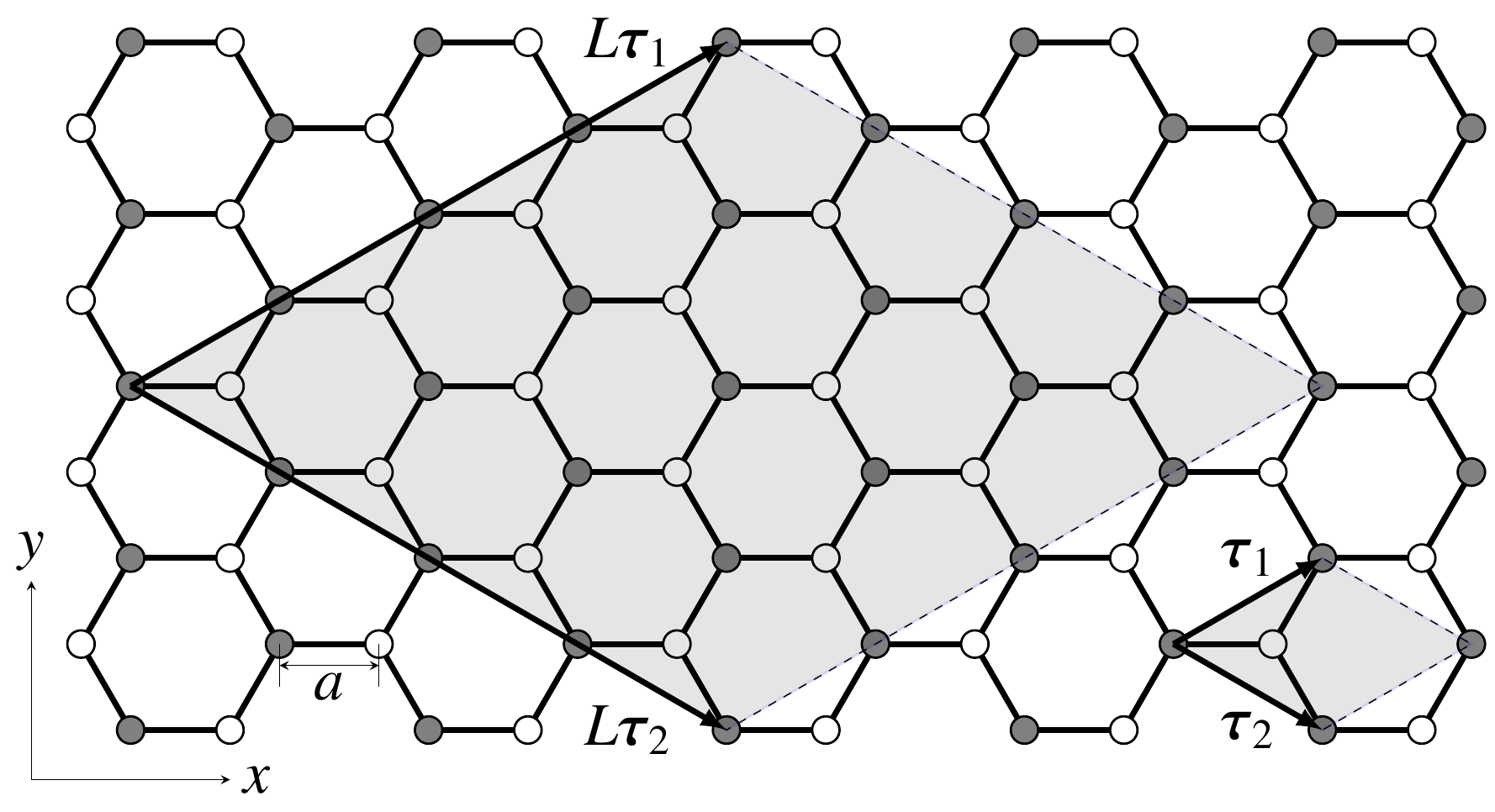}
    \caption{
      \label{fig.cluster} 
      A finite-size cluster and a unit cell of the honeycomb lattice. 
      $\bs{\tau}_1 = a(\frac{3}{2},\frac{\sqrt{3}}{2})$ and
      $\bs{\tau}_2 = a(\frac{3}{2},-\frac{\sqrt{3}}{2})$ are 
      the primitive translational vectors with $a$ being the lattice constant. 
      The $x$ and $y$ axes are indicated in the lower-left part of the figure. 
      The small parallelogram defined by  $\bs{\tau}_1$ and $\bs{\tau}_2$ 
      is the unit cell. 
      The large parallelogram defined by $L\bs{\tau}_1$ and $L\bs{\tau}_2$ 
      is a finite-size cluster of $L=4$. 
      The filled (empty) circles represent lattice sites belonging to sublattice $A$ ($B$). 
    }
  \end{center}
\end{figure}

The Hamiltonian of the Hubbard model on the honeycomb lattice is given by 
\begin{equation}
\hat{H} = \hat{H}_t  + \hat{H}_U, 
\end{equation}
where
\begin{eqnarray}
  \hat{H}_t &=& t \sum_{i} \sum_{\s=\up,\dn} 
  \left( \hat{c}_{A,\bs{r}_i,\s}^\dag \hat{c}_{B,\bs{r}_i,\s} 
         +\hat{c}_{A,\bs{r}_i+\bs{\tau}_1,\s}^\dag \hat{c}_{B,\bs{r}_i,\s} \right. \nonumber \\
         & & \quad\quad\quad\quad\quad\quad \left. +\hat{c}_{A,\bs{r}_i+\bs{\tau}_2,\s}^\dag \hat{c}_{B,\bs{r}_i,\s} 
  + {\rm H. c.} \right)
  \label{Ham0}
\end{eqnarray}
and
\begin{equation}
  \hat{H}_U = U\sum_{i}\sum_{\alpha=A,B} \hat{n}_{\alpha,\bs{r}_i, \up} \hat{n}_{\alpha,\bs{r}_i, \dn}. 
\end{equation}
Here, $\hat{c}_{\alpha,\bs{r}_i, \s}^\dag$ ($\hat{c}_{\alpha, \bs{r}_i, \s}$) is a creation (annihilation) 
operator of a fermion at unit cell $i$, located at $\bs{r}_i=n_i^{(1)}\bs{\tau}_1+n_i^{(2)}\bs{\tau}_2$  
(where $n_i^{(1)}$ and $n_i^{(2)}$ are integer), and 
sublattice $\alpha\,(=A,B)$ with spin $\s\,(=\up,\dn)$, and 
$\hat{n}_{\alpha,\bs{r}_i,\s} = \hat{c}_{\alpha,\bs{r}_i,\s}^\dag \hat{c}_{\alpha,\bs{r}_i,\s}$ (see Fig.~\ref{fig.cluster}). 
$t$ is the hopping integral between the nearest-neighbor sites of the honeycomb lattice 
and $U$ is the strength of the on-site interaction. In this paper, we consider fermion density $n_f=1$, i.e., half filling, 
for which the Dirac points are located exactly at the Fermi level in the noninteracting limit. 

Figure~\ref{fig.cluster} shows the honeycomb lattice 
spanned by primitive translational vectors 
$\bs{\tau}_1 = a(\frac{3}{2},\frac{\sqrt{3}}{2} )$ and
$\bs{\tau}_2 = a(\frac{3}{2},-\frac{\sqrt{3}}{2} )$ with 
$a$ being the lattice constant. 
A finite-size cluster of the linear dimension $L$ is defined by the two vectors 
$L\bs{\tau}_1$ and $L\bs{\tau}_2$, containing $N_{\rm cell}=L^2$ 
unit cells and hence $N_{\rm site}=2N_{\rm cell} = 2L^2$ sites.  
We choose the clusters of 
$L=$ 8, 14, 20, 26, 32, 38, 44, 50, 62, and 74 under periodic boundary conditions, 
for which the closed-shell condition is satisfied~\cite{Sorella_PRB2015}. 
The maximum size considered here thus contains 10952 sites, which is substantially 
(more than four times) larger than the previous largest AFQMC simulations of the
two-dimensional Hubbard models~\cite{Otsuka2016,Sorella2012}.

\subsection{Auxiliary-field quantum Monte Carlo method}

We study the ground-state properties of the Hubbard model $\hat{H}$ 
with the zero-temperature AFQMC method~\cite{Sugiyama1986,Sorella1988,Sorella1989,Becca_Sorella_book}, 
where the ground-state expectation value of an operator $\hat{\mcal{O}}$ is evaluated as 
\begin{equation}
  \left\langle \hat{\mcal{O}} \right\rangle 
  =
  \left\langle \Psi_0 \left| \hat{\mcal{O}} \right| \Psi_0 \right\rangle 
  = 
  \lim_{\tau\to\infty}
  \frac{
    \langle \Psi_{\rm T} | \e^{-\tau \hat{H}/2} \hat{\mathcal{O}}
    \e^{-\tau \hat{H}/2}| \Psi_{\rm T} \rangle }
       {\langle \Psi_{\rm T} | \e^{-\tau \hat{H}}  |\Psi_{\rm T} \rangle }, 
       \label{afqmc}
\end{equation}
where $|\Psi_0 \rangle$ is the normalized ground state of $\hat{H}$, 
$\tau \geqslant 0 $ is the projection time, and $ |\Psi_{\rm T} \rangle $ is a 
trial wavefunction such that $\langle \Psi_0 | \Psi_{\rm T} \rangle \not = 0$. 
We choose as $|\Psi_{\rm T} \rangle$ the ground state of $\hat{H}_t$, i.e., the Fermi sea. 

The imaginary-time evolution is performed 
with the second-order Trotter-Suzuki decomposition  
$
  \e^{-\tau \hat{H}} = \prod_{l=1}^{N_\tau} 
  (
  \e^{-\Delta_\tau \hat{H}_t/2} 
  \e^{-\Delta_\tau \hat{H}_U} 
  \e^{-\Delta_\tau \hat{H}_t/2} 
  ) + O(\Delta_\tau^2), 
$
where $\tau$ is discretized into $N_\tau$ time slices  
with an interval $\Delta_\tau=\tau/N_\tau$ and 
$O(\Delta_\tau^2)$ is the systematic error 
due to the imaginary-time discretization~\cite{Trotter1959,Suzuki1976}. 
At each time slice $l$, the discrete version
of the Hubbard-Stratonovich transformation 
\begin{equation}
  \e^{-\Delta_\tau \hat{H}_U} = 
  C
  \sum_{s_{A,1}} \sum_{s_{B,1}}
  \cdots
  \sum_{s_{B,N_{\rm cell}}}
  \exp{\left[\lambda  \sum_{\alpha,i} s_{\alpha, i} 
      \left(\hat{n}_{\alpha, \bs{r}_i,\up} - \hat{n}_{\alpha, \bs{r}_i,\dn}\right)\right]} 
  \label{hst}
\end{equation} 
is applied, 
where $s_{\alpha, i}=\pm1$ is the auxiliary field on 
sublattice $\alpha$ of the unit cell at $\bs{r}_i$, 
$\cosh(\lambda) = \e^{\Delta_\tau U/2}$,  
and $C=(e^{-\Delta_\tau U/4}/2)^{2L^2}$~\cite{Hirsch1983,Hubbard1959,Stratonovich1957}.  
When this equation is used to evaluate the full propagator
$\prod\limits_l \exp(-\Delta_\tau \hat{H})$, an explicit imaginary-time ($l$) dependence
of the field $s_{\alpha, i}=s_{\alpha, i}(l)$ appears in each time slice,
according to Eq.~(\ref{hst}). 
The multiple summation over $\{s_{\alpha, i}(l)\}$ is performed 
by the Monte Carlo method with the importance sampling.  
The negative-sign problem does not arise at half filling 
owing to the particle-hole symmetry~\cite{Hirsch_PRB1985}. 
In this study, we set $\Delta_\tau t=0.1$ without attempting  
the extrapolation $\Delta_\tau \to 0$ because it already
provides a satisfactory accuracy ($<2\%$) in all correlation functions studied.
Large enough projection times
 $\tau t=50$ or equivalently $N_\tau = 500$
($\tau t=80$ or equivalently $N_\tau = 800$)
for clusters of $L \leqslant 20$ ($L \geqslant 26$)
are used to obtain the converged $\tau \to \infty$ results in Eq.~(\ref{afqmc}).

\subsection{Sparse-matrix exponential}

One of the most computationally expensive operations in the AFQMC method for large clusters 
is the multiplication of
$e^{\pm \Delta_\tau \bs{H}_t}$ (or $e^{\pm \Delta_\tau \bs{H}_t/2}$) 
to the wavefunction matrix or to the  Green's function matrix, where 
$\bs{H}_t$ is the (real-space) matrix representation of $\hat{H}_t$.
Usually, $e^{\pm \Delta_\tau \bs{H}_t}$ is treated as
an $N_{\rm site} \times N_{\rm site}$ 
dense matrix
with the spectral decomposition 
$e^{\pm \Delta_\tau \bs{H}_t}=\bs{U}^Te^{\pm \Delta_\tau \bs{D}} \bs{U}$, 
where 
$\bs{U}$ is a $N_{\rm site} \times N_{\rm site}$ 
orthogonal matrix that diagonalizes $\bs{H}_t$, i.e., $\bs{H}_t \bs{U}=\bs{U}\bs{D}$.
Although $e^{\pm \Delta_\tau \bs{D}}$ is diagonal, $\bs{U}$ is generally dense and thus 
$e^{\pm \Delta_\tau \bs{H}_t}$ is dense. 
Therefore, the computational cost of the matrix-matrix multiplication scales as $O(N_{\rm site}^3)$.
Here, we describe an alternative multiplication scheme of $e^{\pm \Delta_\tau \bs{H}_t}$ 
which is efficient for large clusters by taking full advantage of the sparseness of $\bs{H}_t$.

In this scheme, we expand the matrix exponential as a polynomial of degree $M$, i.e., 
\begin{equation}
  \e^{\pm \Delta_\tau \bs{H}_t}\approx
  {\mathcal I}_0(\rho \Delta_\tau) \bs{I} + 2 \sum_{k=1}^M (\pm1)^k {\mathcal I}_k(\rho\Delta_\tau) T_k (\tilde{\bs{H}}_t), 
  \label{eq.Chebyshev}
\end{equation} 
where $\bs{I}$ is the identity matrix, 
${\mathcal I}_k(\rho \Delta_\tau)$ is the $k$th order modified Bessel function of the first kind,
$\rho$ is the spectral radius of $\bs{H}_{t}$
($\rho=3|t|$ in the present case), and 
$\tilde{\bs{H}}_{t} = \bs{H}_t/\rho$. 
$T_k(\tilde{\bs{H}}_t)$ is the $k$th order Chebyshev polynomial of the first kind, which can be 
obtained iteratively as 
$T_0(\tilde{\bs{H}}_t)=\bs{I}$, $T_1(\tilde{\bs{H}}_t)=\tilde{\bs{H}}_t$,
and $T_k(\tilde{\bs{H}}_t)=2\tilde{\bs{H}}_t T_{k-1}(\tilde{\bs{H}}_t)-T_{k-2}(\tilde{\bs{H}}_t)$ 
for $k\geqslant 2$~\cite{TalEzer1984,Vijay2002,Iitaka2003,Weisse_book}.
A similar orthogonal-polynomial expansion of the Boltzmann factor with the Legendre polynomial has been employed in 
a finite-temperature dynamical density-matrix-renormalization-group method~\cite{Sota2008}.
As shown below, we find that, for large $N_{\rm site}$, 
the multiplication of $\e^{\pm \Delta_\tau \bs{H}_t}$ with manipulating 
$\bs{H}_t$ as a sparse matrix in the right-hand side of~(\ref{eq.Chebyshev})  
is faster than 
the direct multiplication of the dense matrix $ \e^{\pm \Delta_\tau \bs{H}_t}$, 
even when  machine accuracy is reached  with large enough $M$.

\begin{figure}
  \begin{center}
    \includegraphics[width=1.0\columnwidth]{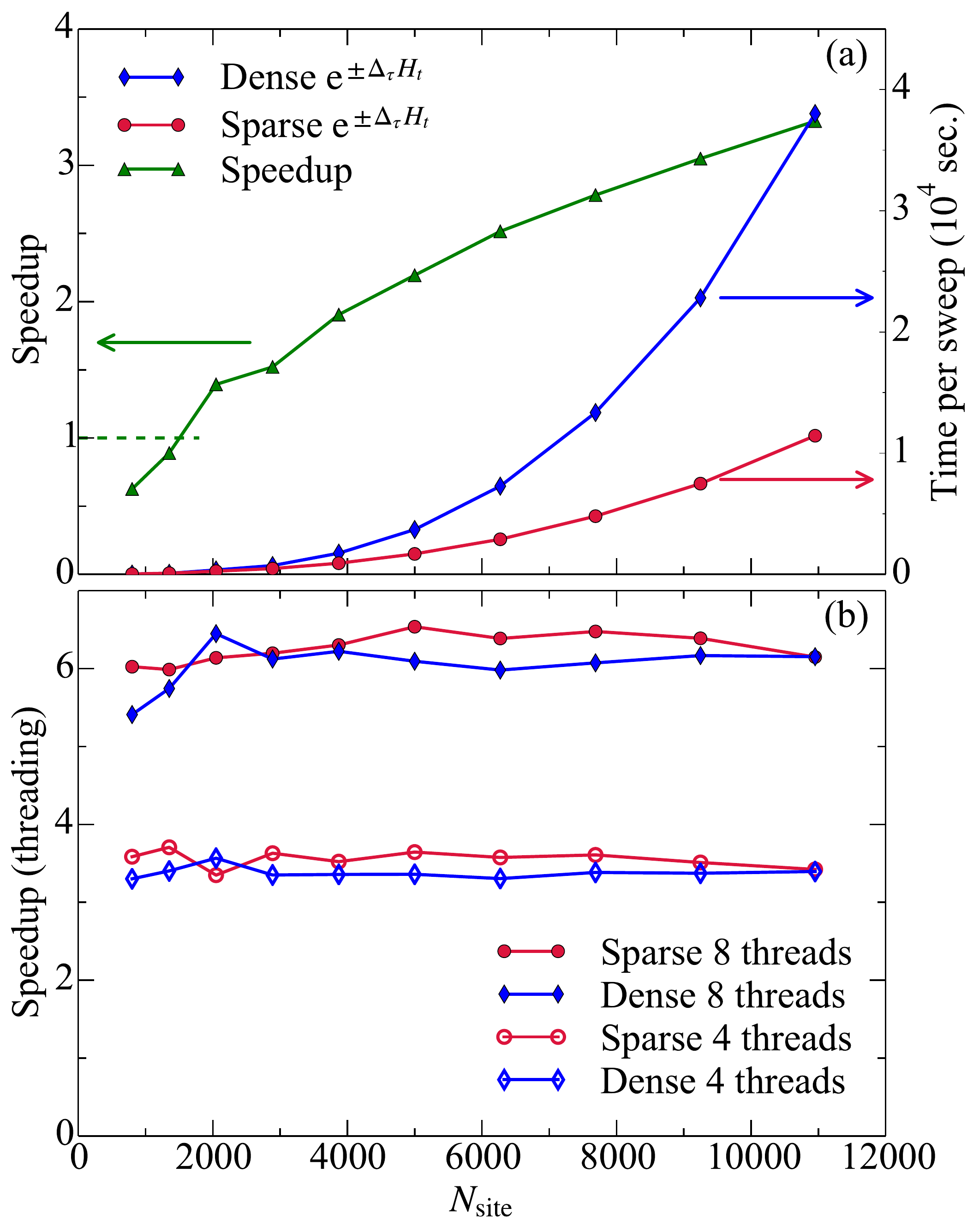}
    \caption{ 
      \label{fig.Chebyshev}
      (a) Computational time of one space-time Monte Carlo sweep 
      with the two multiplication schemes (right axis) 
      and speedup of the sparse-matrix case in the polynomial expansion scheme 
      relative to the dense-matrix case in the conventional scheme (left axis) 
      for $N_{\rm site}=2L^2$ with $L=$ 20, 26, 32, 38, 44, 50, 56, 62, 68, and 74.  
      The dashed line at ${\rm Speedup}=1$ is a guide to the eye. 
      A single thread is used for the calculations.       
      (b) Speedup with multi threading relative to the single-thread case. 
      These benchmark calculations are performed at the HOKUSAI GreatWave facility 
      with SPARC64 XIfx processors in RIKEN. 
    }
  \end{center}
\end{figure}

Figure~\ref{fig.Chebyshev}(a) shows 
the computational time of one space-time Monte Carlo sweep with the two multiplication schemes 
for fixed $N_{\rm \tau}=100$, $\Delta_\tau t=0.1$, and $U/t=3.5$. 
The same initial auxiliary field configuration $\{s_{\alpha,i}(l) \}$ with the same random seed for 
the same random number generator is used for both schemes.
The stabilization (i.e., orthonormalization) of the wavefunction~\cite{Sorella1989,Imada1989,White1989} is made 
every 10 time slices.  
$M=8$ ($M=7$) is used for the expansion with 
$\pm \Delta_\tau t$ ($\Delta_\tau t/2$) to achieve an accuracy of $<10^{-13}$ (see below). 
Since $\bs{H}_t$ has only $z_{\rm c}$ 
($z_{\rm c}$: the coordination number, i.e., $z_{\rm c}=3$ for the honeycomb lattice) 
nonzero matrix elements in each column and row (thus, totally $z_{\rm c} N_{\rm site}$ nonzero elements), 
the computational cost of the multiplication of $e^{\pm \Delta_\tau \bs{H}_t}$ to 
an $N_{\rm site} \times N_{\rm site}$ dense matrix scales as $O(z_{\rm c} M N_{\rm site}^2)$ 
when the polynomial expansion scheme in Eq.~(\ref{eq.Chebyshev}) is employed.    
A convenient speedup larger than one is achieved for $N_{\rm site}\gtrsim 2000$ in our computing environment 
and increases with $N_{\rm site}$. 
Modern processors have the possibility to perform several independent tasks,
called ''threads'' within the same computational unit. 
As shown in Fig.~\ref{fig.Chebyshev}(b), 
the speedup with threading is also as effective as 
that in the dense-matrix case. 
Here, the compressed-row-storage (CRS) format 
(see for example Ref.~\cite{Hager_book}) is used 
to store the nonzero matrix elements of $\bs{H}_t$ 
within the polynomial expansion scheme.

In analogy with the high-temperature series expansion~\cite{Imada1986,Jaklic2000}, 
the convergence of the polynomial expansion in Eq.~(\ref{eq.Chebyshev}) with relatively small $M$
is evident 
because usually $\Delta_\tau$ is taken small ($\Delta_\tau t \ll 1$) in the AFQMC simulation. 
Given a desired accuracy $\epsilon$ for the polynomial expansion, 
$M$ can be determined to satisfy 
\begin{equation}
  \left|\left|
  \e^{\pm \Delta_\tau \bs{H}_t} - 
  \left(
    {\mathcal I}_0(\rho \Delta_\tau) \bs{I} + 2 \sum_{k=1}^M (\pm1)^k {\mathcal I}_k(\rho\Delta_\tau) T_k (\tilde{\bs{H}}_t)
    \right)
  \right| \right|_{\rm max}  < \eps,
  \label{truncation}
\end{equation}
where $||\bs{A}||_{\rm max} = \max_{ij} |A_{ij}|$ is 
the maximum norm of $\bs{A}$ and $A_{ij} = [\bs{A}]_{ij}$. 
We set $\eps =10^{-13}$ 
and find that $M=8$ ($M=7$) is the minimum value that satisfies
the inequality in Eq.~(\ref{truncation})
for $\Delta_\tau t =0.1$ ($\Delta_\tau t/2 =0.05$ when $e^{\pm \Delta_\tau \bs{H}_t/2}$ is expanded), 
irrespectively of the system size.  
This implies that the polynomial expansion is well controlled and 
even does not introduce the additional systematic error 
by terminating the expansion at finite $M$ as $\epsilon$ is 
negligibly smaller than the statistical error.
Finally, we note that, 
if the degree $M$ is the same,
the Chebyshev polynomial expansion in Eq.~(\ref{eq.Chebyshev}) gives 
better accuracy than the Taylor expansion
$\e^{\pm \Delta_\tau \bs{H}_t} \approx \sum_{k=0}^{M} \frac{(\pm\Delta_\tau)^k}{k!} \bs{H}_t^k$, 
in the sense that the matrix norm of the difference from the exact $\e^{\pm \Delta_\tau \bs{H}_t}$ is smaller, 
for the model studied here.

Recently, a different approach
to reduce the computational effort of fermionic quantum Monte Carlo (QMC) simulations,  
dubbed as effective momentum ultra-size QMC,  
has been proposed and successfully used in some model systems~\cite{Liu2017,Liu2018}. 
This approach is designed to capture the low-energy physics for original lattice models of interest.

\section{Quasiparticle weight}\label{sec3}
The main quantity considered here is the equal-time single-particle Green's function 
\begin{eqnarray}
  D_{AB,\s}(\bs{r}) 
  &=& 
  \frac{1}{N_{\mathrm{cell}}}\sum_{\bs{r}^{\prime}}
  \left \langle \hat{c}_{B,\bs{r}'+\bs{r},\s}^\dag \hat{c}_{A,\bs{r}',\s} \right \rangle,  
  \label{DAB0}
\end{eqnarray}
where 
$\bs{r}$ denotes a relative spatial position of two unit cells at $\bs{r}'$ and $\bs{r}'+\bs{r}$,  
and the average 
$ \langle \cdots \rangle = 
{{\rm Tr}[\e^{-\hat{H}/T} \cdots]}/
{{\rm Tr}[\e^{-\hat{H}/T}]} $
is defined at a finite temperature $T$ 
for the clarity of the following formulation. 
The zero-temperature limit will be taken only at the end of the calculation. 
Since $D_{AB,\s}(\bs{r})$ represents the probability amplitude 
 that a hole created 
on sublattice $A$ in the unit cell at $\bs{r}'$ propagates
to sublattice $B$ in the unit cell at $\bs{r}+\bs{r}'$, 
the long-distance behavior of $D_{AB,\s}(\bs{r})$ should enable us to 
distinguish whether the system is semimetallic or insulating. 
Indeed, as shown in Appendixes~\ref{appA} and \ref{appB}, $D_{AB,\s}(\bs{r})$ decays 
algebraically, with a prefactor proportional to $Z$, in the semimetallic phase, while 
it decays exponentially in the insulating phase.

\subsection{Noninteracting limit}

First, we analyze $D_{AB,\s}(\bs{r})$ in the noninteracting limit. 
For this purpose, we diagonalize $\hat{H}_t$ as 
\begin{eqnarray}
  \hat{H}_t 
  &=& \sum_{\bs{k},\s} \left(
   |h_{\bs{k}}|  \hat{\psi}_{+,\bs{k},\s}^\dag \hat{\psi}_{+,\bs{k},\s}
  -|h_{\bs{k}}|  \hat{\psi}_{-,\bs{k},\s}^\dag \hat{\psi}_{-,\bs{k},\s}
  \right),
  \label{H0}
\end{eqnarray}
where 
$h_{\bs{k}}= t\left(1 + \e^{-\imag \bs{k}\cdot\bs{\tau}_1} + \e^{-\imag \bs{k}\cdot\bs{\tau}_2}\right)$, 
$
\hat{\psi}_{+,\bs{k},\s} = 
\frac{1}{\sqrt{2}} \left( \hat{c}_{A,\bs{k},\s} + \e^{\imag \theta_{\bs{k}}} \hat{c}_{B,\bs{k},\s} \right) 
$, 
$
\hat{\psi}_{-,\bs{k},\s} = 
\frac{1}{\sqrt{2}} \left( \hat{c}_{A,\bs{k},\s} - \e^{\imag \theta_{\bs{k}}} \hat{c}_{B,\bs{k},\s} \right)  
$, 
$\e^{\imag \theta_{\bs{k}}} = h_{\bs{k}}/|h_{\bs{k}}|$, 
and 
$\hat{c}_{\alpha,\bs{k},\s}=N_{\rm cell}^{-1/2}\sum_{i}  \hat{c}_{\alpha,\bs{r}_i,\s} \e^{-\imag \bs{k}\cdot\bs{r}_i} $.  
The bonding- and antibonding-band energies are $-|h_{\bs{k}}|$ and $|h_{\bs{k}}|$, respectively.  
The zero-energy modes protected by the chiral symmetry~\cite{Semenoff2012,Hatsugai2013} 
appear at two inequivalent momenta, $K$ and $K'$ points, 
which are specified by the vectors 
$\bs{k}_K    = \frac{1}{a}(\frac{2\pi}{3},\frac{2\pi}{3\sqrt{3}})$ and 
$\bs{k}_{K'} = \frac{1}{a}(\frac{2\pi}{3},-\frac{2\pi}{3\sqrt{3}})$, respectively.  
$D_{AB,\s}(\bs{r})$ in the noninteracting limit is now evaluated as   
\begin{eqnarray}
  &&D_{AB,\s}^{(0)}(\bs{r}) 
  = \ksum \left \langle \hat{c}_{B,\bs{k},\s}^\dag \hat{c}_{A,\bs{k},\s} \right \rangle \e^{\imag \bs{k}\cdot \bs{r}} \notag \\
  &=& \ksumhalf \left[n_{\rm F}(|h_{\bs{k}}|) - n_{\rm F}(-|h_{\bs{k}}|)\right] 
  \frac{h_{\bs{k}}}{|h_{\bs{k}}|} \e^{\imag \bs{k}\cdot \bs{r}} \label{D0sum}  \\
  &\underset{T\to0}{=}& -\frac{1}{2N_{\rm cell}} \sum_{\bs{k}\not = \bs{k}_K, \bs{k}_{K'}}
  \frac{h_{\bs{k}}}{|h_{\bs{k}}|} \e^{\imag \bs{k}\cdot \bs{r}},   
  \label{D0sum0}
\end{eqnarray}
where the superscript ``$(0)$'' denotes that the quantity is in the noninteracting limit. 
$n_{\rm F}(E) = 1/(\e^{E/T} + 1)$ is the Fermi distribution function, which arises from 
the occupation of the fermions 
$\left\langle \hat{\psi}^\dag_{\pm,\bs{k},\s} \hat{\psi}_{\pm,\bs{k},\s} \right\rangle = n_{\rm F}(\pm|h_{\bs{k}}|)$.  
The summand in Eq.~(\ref{D0sum}) exactly at the $K$ and $K'$ points 
is zero because $n_{\rm F}(|h_{\bs{k}_{K(K')}}|) - n_{\rm F}(-|h_{\bs{k}_{K(K')}}|)=0$ 
and thereby these two momenta are excluded from the summation in Eq.~(\ref{D0sum0}).

We should note that, on the contrary to $D_{AB,\s}^{(0)}(\bs{r})$, $D_{AA,\s}^{(0)}(\bs{r})$ at half filling 
gives merely a trivial $\bs{r}$ dependence, i.e.,  
\begin{equation}
  D_{AA,\s}^{(0)}(\bs{r}) 
= \frac{1}{N_{\rm cell}} \sum_{\bs{k}} 
\left\langle \hat{c}_{A,\bs{k},\s}^\dag \hat{c}_{A,\bs{k},\s} \right\rangle
\e^{\imag \bs{k}\cdot\bs{r}} 
= \frac{1}{2} \delta_{\bs{r},\bs{0}},  
\end{equation}
because $\left\langle \hat{c}_{A,\bs{k},\s}^\dag \hat{c}_{A,\bs{k},\s}\right\rangle=1/2$. 
Here, $ \delta_{\bs{r},\bs{0}}=1$ when $\bs{r}=\bs{0}$ and zero otherwise. 
This is also the case when the interaction $U$ is finite because  
$\langle \hat{c}_{A\bs{k}\s}^\dag \hat{c}_{A\bs{k}\s}\rangle=1/2$ as long as the particle-hole symmetry is preserved.  
Therefore, $D_{AA,\s}(\bs{r})$ and similarly $D_{BB,\s}(\bs{r})$ do not 
show any long-distance propagation of a hole that can discriminate the nature of the different ground states.

\subsection{Interacting case}

In order to analyze $D_{AB,\s}(\bs{r})$ in an interacting system, 
we now express this quantity with the single-particle Green's function $G_{AB,\s}(\bs{r},\imag \w_\nu)$ 
in the Matsubara-frequency representation~\cite{Ezawa1957,Matsubara1955},
i.e., 
\begin{eqnarray}
  D_{AB,\s}(\bs{r})
  &=& T \sum_{\nu=-\infty}^{\infty} G_{AB,\s}(\bs{r},\imag \w_\nu) \notag \\
  &=& \ksum \oint_{\mcal{C}} \frac{\dd z}{2\pi\imag} n_{\rm F}(z) G_{AB,\s}(\bs{k},z) \e^{\imag \bs{k}\cdot{\bs{r}}}, 
  \label{DAB1}
\end{eqnarray}
where 
$\imag \w_{\nu} = (2\nu+1)\pi\imag T$ with $\nu$ integer is the fermionic Matsubara frequency, 
$G_{AB,\s}(\bs{r},\imag \w_\nu) = N_{\rm cell}^{-1} \sum_{\bs{k}} G_{AB,\s}(\bs{k},\imag \w_\nu) \e^{\imag \bs{k}\cdot{\bs{r}}}$, 
and the frequency sum is converted to the contour integral. 
The contour $\mcal{C}$ is chosen so as to include all the singularities 
of $G_{AB,\s}(\bs{k},z)$, which lie on the real axis, and therefore does not enclose 
the Matsubara frequencies. 

\begin{figure*}
  \begin{center}
    \includegraphics[width=1.\textwidth]{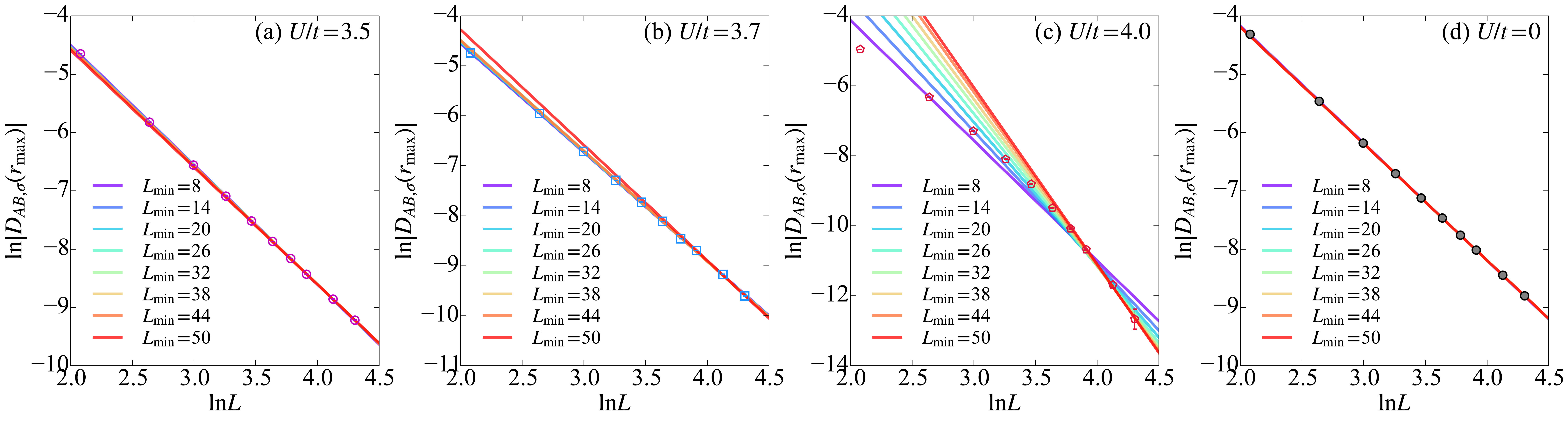}    
    \caption{
      \label{fig.slope}       
      $\ln L$ dependence of $\ln \left|D_{AB,\s}(r_{\rm max})\right|$ 
      for (a) $U/t=3.5$, (b) $U/t=3.7$, and (c) $U/t=4$. 
      For comparison, the result for the noninteracting case is also shown 
      in (d).
      Lines are linear fit to the data of the form $-\alpha(L_{\rm min}) \ln L + b(L_{\rm min})$, 
      where $\alpha(L_{\rm min})$ and $b(L_{\rm min})$ are fitting parameters 
      with $L_{\rm min}$ being the minimum $L$ used for the fit. 
      The maximum $L$ used for the fit is 74 for all cases, including (d). 
    }
  \end{center}
\end{figure*}

We now assume that the single-particle Green's function 
near the Fermi level has a Fermi-liquid-type pole~\cite{AGD}, 
which should be consistent with the particle-hole symmetry of the model, i.e., 
\begin{eqnarray}
  \bs{G}_{\s}(\bs{k},z) 
  &=&
  \left[
    \begin{array}{cc}
      G_{AA,\s}(\bs{k},z)  & G_{AB,\s}(\bs{k},z)  \\ 
      G_{BA,\s}(\bs{k},z)  & G_{BB,\s}(\bs{k},z)  
    \end{array}
    \right] \nonumber \\
  &=& 
  \frac{Z}{z^2-|\tilde{h}_{\bs{k}}|^2}
  \left[
    \begin{array}{cc}
    z & \tilde{h}_{\bs{k}}\\
    \tilde{h}_{\bs{k}}^* & z 
    \end{array}
    \right] 
  + { \rm (incoherent~part)},
  \label{Gint}
\end{eqnarray}
where $\tilde{h}_{\bs{k}}=(v_{\rm F}/v_{\rm F}^{(0)}) h_{\bs{k}}$ 
with $v_{\rm F}$ and $v_{\rm F}^{(0)}(=3|t|a/2)$ 
being the Fermi velocity of the interacting and noninteracting systems, respectively, and 
$Z$ is the quasiparticle weight 
at the nodal Dirac point.
The incoherent part is a function of $z$ and the singularities lie well away from 
the Fermi level. 

By substituting $G_{AB,\s}(\bs{k},z)$ of Eq.~(\ref{Gint}) 
into Eq.~(\ref{DAB1}) and  performing the contour integral,  
we obtain in the large-distance limit ($|\bs{r}|/a \gg 1$) that
\begin{eqnarray}
  D_{AB,\s}(\bs{r})
  &\approx&  
  \ksumZhalf
  \left[ n_{\rm F}(|\tilde{h}_{\bs{k}}|) - n_{\rm F}(-|\tilde{h}_{\bs{k}}|)\right] 
  \frac{\tilde{h}_{\bs{k}}}{|\tilde{h}_{\bs{k}}|} 
  \e^{\imag \bs{k}\cdot \bs{r}} 
  \notag 
  \\
  &\underset{T\to0}{=}&
  -
  \frac{Z}{2N_{\rm cell}} \sum_{\bs{k}\not = \bs{k}_K, \bs{k}_{K'}}
  \frac{h_{\bs{k}}}{|h_{\bs{k}}|} \e^{\imag \bs{k}\cdot \bs{r}}  
  =
  Z D_{AB,\s}^{(0)}(\bs{r}).    
  \label{DABFL}
\end{eqnarray} 
Here, the incoherent part does not contribute to $D_{AB,\sigma}(\bs{r})$ in the long-distance limit. 
This is because the singularities of the incoherent part appear away from the Fermi level 
and thus the contribution of the incoherent part to $D_{AB,\sigma}(\bs{r})$ decays exponentially in $|\bs{r}|$ 
(see Appendix~\ref{appB}). 
Note that the $K$ and $K'$ points are excluded from the summation in Eq.~(\ref{DABFL}), 
as in the noninteracting case. 
This justifies the use of finite-size clusters with $L=3n+2$ (or $L=3n+1$, where $n$ is integer) 
for our AFQMC simulations, where the closed-shell condition 
in the noninteracting limit is convenient for accurate simulations~\cite{Sorella_PRB2015}.  

The form of $D_{AB,\s}(\bs{r})$ in Eq.~(\ref{DABFL}) is quite natural 
as it matches the simple substitution of the quasiparticle operators 
$\hat{c}_{B,\bs{r},\s}^\dag \mapsto \hat{q}_{B,\bs{r},\s}^\dag = \sqrt{Z} \hat{c}_{B,\bs{r},\s}^\dag$ and 
$\hat{c}_{A,\bs{r},\s} \mapsto \hat{q}_{A,\bs{r},\s} = \sqrt{Z} \hat{c}_{A,\bs{r},\s}$ 
into $D_{AB,\s}^{(0)}(\bs{r})$~\cite{Fabrizio2007}.  
The quasiparticle weight $Z$ at the Fermi point in the thermodynamic limit 
is now simply evaluated via the ratio of the equal-time single-particle Green's functions in the long-distance 
limit, i.e., 
\begin{equation}
  Z =\lim_{|\bs{r}| \to \infty} \frac{D_{AB,\s}(\bs{r})}{D^{(0)}_{AB,\s}(\bs{r})}.  
  \label{ratio}
\end{equation} 
Since the Fermi velocity $v_{\rm F}$, another unknown quantity, 
does not appear here, $Z$ can be estimated independently of $v_{\rm F}$.

\section{Numerical results}\label{sec4}
\begin{figure}
  \begin{center}
    \includegraphics[width=1.\columnwidth]{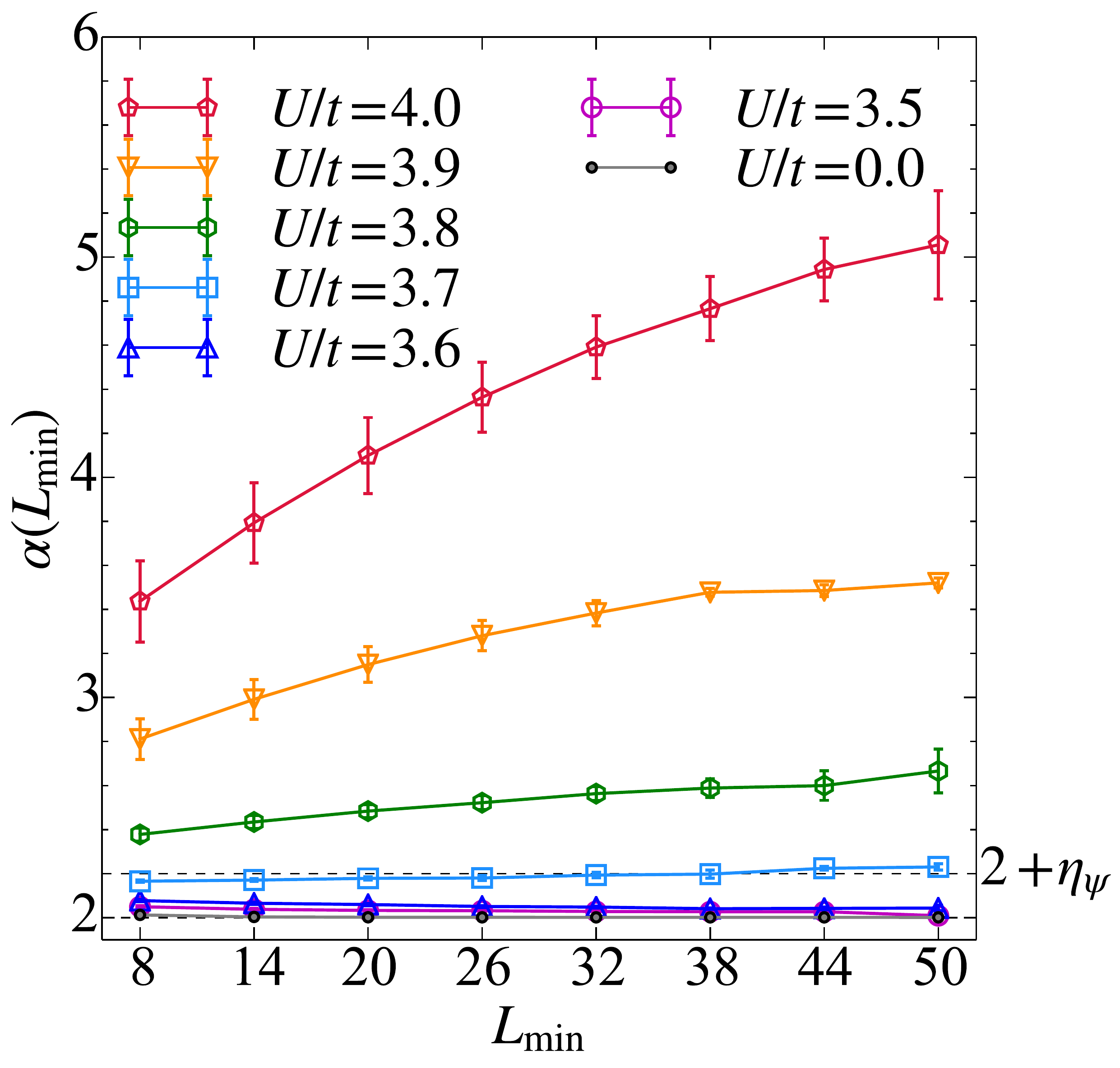}    
    \caption{
      \label{fig.slope2}       
      $L_{\rm min}$ dependence of $\alpha(L_{\rm min})$ for different values of $U$ indicated in the figure.
      For comparison, $\alpha(L_{\rm min})$ for $U=0$ is also shown by grey dots. 
      The dashed lines indicate $\alpha=2$ and $\alpha=2+\eta_{\psi}$ with $\eta_{\psi}=0.2$. 
    }
  \end{center}
\end{figure}

Employing the AFQMC method, 
we now examine numerically 
the long-distance behavior of $D_{AB,\s}(\bs{r})$. 
As shown in Appendixes~\ref{appA} and \ref{appB}, 
$D_{AB,\s}(\bs{r})$ decays in $r=|\bs{r}|$ as 
\begin{equation}
  D_{AB,\s}(r) \sim \frac{1}{r^2} 
\end{equation}
in the Fermi liquid, while $D_{AB,\s}(r)$ decays exponentially in the insulating state. 
Figure~\ref{fig.slope} shows the cluster-size ($L$)
dependence of $D_{AB,\s}(r_{\rm max})$ 
for $U/t=$3.5, 3.7, and 4, where 
$r_{\rm max} = |\bs{r}_{\rm max}|$ is the maximum distance available 
in a given finite-size cluster of linear dimension $L$ (see Fig.~\ref{fig.cluster}). 
We take $\bs{r}_{\rm max}$ in the $x$ direction  
to remove the phase factors in $D_{AB,\s}(r)$ 
(for details, see Appendix~\ref{appA}). 
The lines are linear fits to the data of the form 
$-\alpha(L_{\rm min}) \ln L + b(L_{\rm min})$, where
$\alpha(L_{\rm min})$ and $b(L_{\rm min})$ are fitting parameters
with $L_{\rm min}$ being the minimum $L$ used for the fit. 
As summarized in Fig.~\ref{fig.slope2}, 
$\alpha(L_{\rm min})$ approaches to $2$ for $U/t=3.5$ and $3.6$, as expected for the Fermi liquid, 
while $\alpha(L_{\rm min})$ increases with $L_{\rm min}$ for $U/t=3.8$, $3.9$ and $4$, 
indicating the insulating behavior.
Only in the vicinity of the quantum critical point $U_{\mathrm{c}}/t \simeq 3.7$
separating the semimetal and the antiferromagnetic insulator~\cite{Sorella2012,Otsuka2016},
we observe the non-Fermi-liquid behavior characterized by the
non-trivial exponent of $2+\eta_{\psi}$, where $\eta_{\psi} \simeq 0.2$~\cite{Otsuka2016} is 
the fermion anomalous dimension~\footnote
{
  Note that the value of the fermion anomalous dimension $\eta_{\psi}$
  (as well as other critical exponents) is still controversial
  as it ranges from $0.071(2)$ to $0.242$, depending on  
  numerical and analytical techniques used~\cite{Janssen2014,Otsuka2016,Zerf2017,Knorr2018}
}. 
Therefore, these results already imply 
that the semimetallic phase is the Fermi liquid.

\begin{figure}
  \begin{center}
    \includegraphics[width=1.\columnwidth]{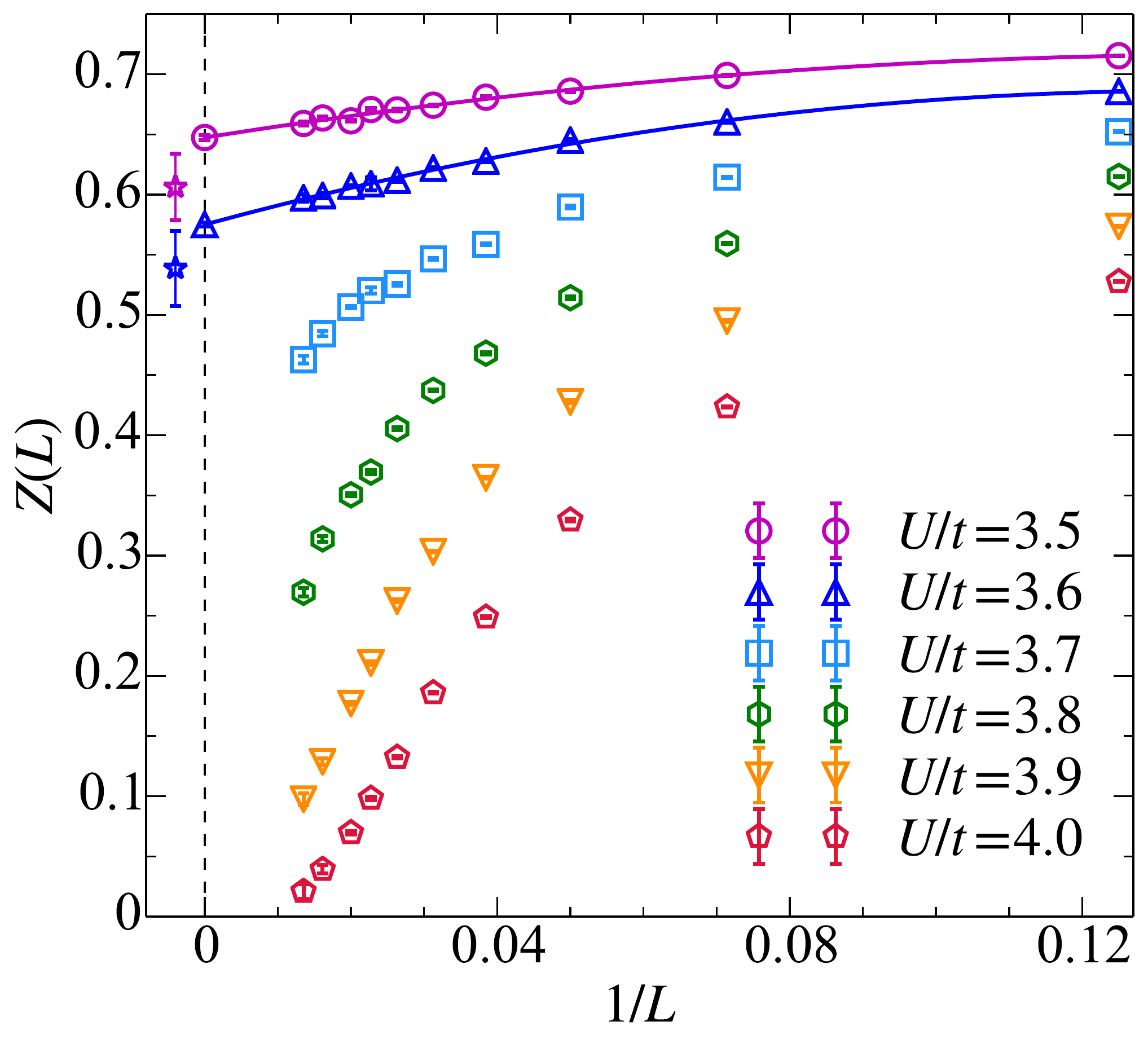}    
    \caption{
      \label{fig.extrapolation} 
      The quasiparticle weight $Z(L)$ given in Eq.~(\ref{ratioL}) 
      as a function of $1/L$. Lines are polynomial fits to the data for $U/t=3.5$ and 3.6.  
      The extrapolated values in the thermodynamic limit are also shown at $1/L=0$. 
      The quasiparticle weight estimated previously from the jump of the momentum distribution 
      function~\cite{Otsuka2016} are also shown by stars next to the present results.
    }
  \end{center}
\end{figure}

Next, we evaluate the quasiparticle weight on finite-size clusters,  
\begin{equation}
  Z(L) = \frac{D_{AB,\s}(\bs{r}_{\rm max})}{D^{(0)}_{AB,\s}(\bs{r}_{\rm max})},  
  \label{ratioL}
\end{equation}
as recently applied by the authors to identify the semimetallic state 
on a triangular lattice~\cite{Otsuka2018}. 
For the Fermi-liquid ground state, 
the quasiparticle weight in the thermodynamic limit, i.e., $Z=\lim_{L\to \infty}Z(L)$, is finite. 
Figure~\ref{fig.extrapolation} shows 
$Z(L)$ as a function of $1/L$ and  
lines are second-order polynomial fits of the form
$\sum_{n=0}^2 c_n L^{-n}$
to the data with $\{c_n\}$ being fitting parameters determined by the least-squares method. 
The extrapolated values of $c_0=Z$ and their error bars in 
the thermodynamic limit are also shown 
at $1/L = 0$ for the semimetallic phase where the Fermi-liquid-like asymptotic behavior is observed in 
$D_{AB,\s}(r)$ (see Fig.~\ref{fig.slope2}). 
We find that these extrapolated values are consistent,  
within two standard deviations,
with our previous results~\cite{Otsuka2016} which are 
estimated from the jump of the momentum distribution function and 
indicated by stars in Fig.~\ref{fig.extrapolation}.  
Our new calculations with Eq.~(\ref{ratioL}) performed on the larger clusters 
are however more accurate as the error bars are more than six-times smaller, 
supporting the validity of the Fermi-liquid theory
in the semimetallic phase of the Honeycomb lattice. 

\section{Conclusions and discussions}\label{sec5}
In conclusion, we have shown by the AFQMC method that
a Fermi-liquid ground state is realized 
in the semimetallic phase of the Hubbard model on the honeycomb lattice at half filling.
This conclusion is obtained by studying the asymptotic behavior of the equal-time single-particle 
Green's function $D_{AB,\s}(r)\sim1/r^2$ and by providing firm numerical indication of
a finite quasiparticle weight $Z$ in the semimetallic phase. 
The finite $Z$ immediately implies the presence of the quasiparticles, each of which 
carries a spin $1\over 2$ and a charge $-e$ (for many electron systems) with the Fermi surface 
unaltered from the noninteracting one, due to  the particle-hole symmetry~\cite{Luttinger1960,Seki2017}. 
In the vicinity of the quantum critical point, 
the non Fermi liquid behavior characterized with
a non-trivial exponent is also probed  directly by the asymptotic behavior of $D_{AB,\s}(r)$. 

Considering the Hubbard model as the minimal model for graphene~\cite{Schuler2013}, 
our results imply a realization of Fermi liquid in graphene, which 
has been often assumed, for example, in Ref.~\cite{Katsnelson2008}.  
However, because of the vanishing 
density of states at half filling, 
the unscreened
long-range Coulomb interactions 
are certainly important for a more realistic modeling of
graphene to examine a possible non-Fermi-liquid behavior 
accompanied with the diverging Fermi 
velocity~\cite{Gonzalez1999,Kotov2012,Ulybyshev2013,Wu2014,Tang2015,Tupitsyn2017,Tang2018,Buividovich2018}. 
Indeed, an anomalous increase of the Fermi velocity in graphene 
has been reported experimentally~\cite{Elias2011}. 
The Hubbard-type models with long-range Coulomb interaction~\cite{Hohenadler2014}  
on the honeycomb lattice might be promising to investigate the non-Fermi-liquid state in graphene 
and also other possible many-body electronic states 
in carbon-based low-dimensional materials such as condensed excitonic states~\cite{Phan2012,Varsano2017}.

\acknowledgements
We acknowledge Tomonori Shirakawa for useful discussions. 
This work has been supported in part by 
Grant-in-Aid for Scientific Research from MEXT Japan 
(under Grant Nos. 26400413 and 18K03475), 
RIKEN iTHES Project, and 
the Simons Collaboration on the Many Electron Problem. 
The numerical simulations have been performed 
on the HOKUSAI supercomputer at RIKEN 
(Projects No.~G17030, No.~G17032, No.~G18007, and No.~G18025) 
and 
on the K computer at RIKEN Center for Computational Science (R-CCS) 
through the HPCI System Research Project 
(Projects No.~hp160159, No.~hp170079, No.~hp170162, No.~hp170308, No.~hp170328, and No.~hp180098). 
K. S. acknowledges support from the JSPS Overseas Research Fellowships.

\appendix

\section{
$D_{AB,\s}(\bs{r})$ in the semimetallic phase}\label{appA}

In this Appendix, we show that $D_{AB}(\bs{r})$ decays 
algebraically in $|\bs{r}|$ for $|\bs{r}|/a\gg 1$ in the semimetallic phase. 
First, we consider the noninteracting limit. 
To examine the asymptotic form of $D^{(0)}_{AB,\s}(\bs{r})$, 
we replace the sum over discrete $\bs{k}$ in Eq.~(\ref{D0sum0}) 
by the integral over continuous $\bs{k}$ in the whole first Brillouin zone, i.e.,  
\begin{equation}
  \ksum \cdots \to \kint \cdots, 
\end{equation}
where $S_{\rm cell}  = 3\sqrt{3}a^2/2$ is the area of the unit cell.  
This is justified in the thermodynamic limit and useful for analyzing the low-energy and long-distance behavior. 
In the thermodynamic limit, Eq.~(\ref{D0sum0}) now reduces to 
\begin{eqnarray}
  D_{AB,\s}^{(0)}(\bs{r}) 
  &=& -\frac{1}{2} \kint \frac{h_{\bs{k}}}{|h_{\bs{k}}|} \e^{\imag \bs{k}\cdot \bs{r}}.
\end{eqnarray}

Since the long-distance behavior of $D_{AB,\s}^{(0)}(\bs{r})$ is dominated by 
the low-energy spectrum around the Dirac ($K$ and $K'$) points, 
we measure momentum $\bs{k}$ from the Dirac points ($v=K,K'$) as 
\begin{equation}
  \bs{k} = \bs{k}_{v} + \bs{q}.
\end{equation}
Expanding $ h_{\bs{k}} = h_{\bs{k}_{K} + \bs{q}}$ around the $K$ point 
with respect to $\bs{q}=(q_x,q_y)$ and taking up to the linear term in $\bs{q}$ yield  
\begin{eqnarray}
  h_{\bs{k}_{K}+\bs{q}} 
  &=&
  t\left(1 
    +\e^{-\imag \frac{4\pi}{3}} \e^{-\imag \bs{q}\cdot\bs{\tau}_1}
    +\e^{-\imag \frac{2\pi}{3}} \e^{-\imag \bs{q}\cdot\bs{\tau}_2}
    \right)
    \notag \\
  &\simeq& t
  \left[1 
    +\e^{-\imag \frac{4\pi}{3}} \left(1-\imag \bs{q}\cdot\bs{\tau}_1\right)
    +\e^{-\imag \frac{2\pi}{3}} \left(1-\imag \bs{q}\cdot\bs{\tau}_2\right)
    \right] \notag \\
  &=&
  \frac{3ta}{2}
  \left(\imag q_x + q_y \right). 
  \label{linearK} 
\end{eqnarray} 
The contribution to $D_{AB,\s}^{(0)}(\bs{r})$ 
from the momentum around the $K$ point is thus evaluated as  
\begin{eqnarray}
  &&-\frac{1}{2} \frac{S_{\rm cell}}{(2\pi)^2} \e^{\imag \bs{k}_{K} \cdot \bs{r}}\int \dd^2 \bs{q} 
  \frac{\imag q_x + q_y}{q} \e^{\imag \bs{q}\cdot \bs{r}} \notag \\
&=&-\frac{1}{2} \frac{S_{\rm cell}}{(2\pi)^2} \e^{\imag \bs{k}_{K} \cdot \bs{r}} 
  \left(\frac{\partial}{\partial r_x} - \imag \frac{\partial}{\partial r_y} \right)
  \int \dd^2 \bs{q}  \frac{1}{q} \e^{\imag \bs{q}\cdot \bs{r}}\notag \\
&=&-\frac{S_{\rm cell}}{4\pi} 
  \e^{\imag \bs{k}_K \cdot \bs{r}} \frac{r_x - \imag r_y}{r^3}, 
  \label{fromK}
\end{eqnarray}
where $q =|\bs{q}|$, $\bs{r}=(r_x,r_y)$, and $r=|\bs{r}|$. 
Here, the integral in the second line is treated as    
\begin{equation}
  \frac{1}{2\pi}\int_0^{\Lambda} \dd q \int_0^{2\pi} \dd \phi \e^{\imag q r \cos{\phi}}
  =  \frac{1}{r} \int_0^{r \Lambda} \dd s J_0(s), 
  \label{Bessel}
\end{equation}
where $s=qr$, $J_0(s)$ is the zeroth-order Bessel function of the first kind,  
and $\Lambda$ is a cutoff momentum of order $\Lambda \sim 1/a$.  
The upper bound of the integral satisfies $r \Lambda \gg 1$ 
because our interest is in the long-distance ($r/a \gg 1$) behavior.   
Since the long-distance behavior of the hole propagation should not 
be affected by 
the cutoff momentum $\Lambda$, 
it is possible to set $r \Lambda \to \infty$. 
Then, the integral of the Bessel function can be performed as $\int_0^\infty \dd s J_0(s)=1$  
and Eq.~(\ref{Bessel}) results in $1/r$, 
as in the Fourier transform (or the Hankel transform) 
of the Coulomb potential in two dimensions
\begin{equation}
  \frac{1}{2\pi}\int \dd^2 \bs{q} \frac{1}{q}\e^{\imag \bs{q}\cdot\bs{r}} = \frac{1}{r}. 
  \label{Coulomb}
\end{equation}
Therefore, the propagation of a hole is long ranged.

Similarly, around the $K'$ point, $h_{\bs{k}_{K} + \bs{q}}$ can be expanded as 
\begin{eqnarray}
  h_{\bs{k}_{K'}+\bs{q}} 
  &=&
  t\left(1 
  +\e^{-\imag \frac{2\pi}{3}} \e^{-\imag \bs{q}\cdot\bs{\tau}_1}
  +\e^{-\imag \frac{4\pi}{3}} \e^{-\imag \bs{q}\cdot\bs{\tau}_2} 
  \right)
  \notag \\
  &\simeq& t
  \left[1 
    +\e^{-\imag \frac{2\pi}{3}} \left(1-\imag \bs{q}\cdot\bs{\tau}_1\right)
    +\e^{-\imag \frac{4\pi}{3}} \left(1-\imag \bs{q}\cdot\bs{\tau}_2\right)
    \right] \notag \\
  &=&
  \frac{3ta}{2}
  \left(\imag q_x - q_y \right).  
  \label{linearK'}
\end{eqnarray}
The contribution to $D_{AB,\s}^{(0)}(\bs{r})$ 
from the momentum around the $K'$ point is thus evaluated as  
\begin{eqnarray}
  &&-\frac{1}{2} \frac{S_{\rm cell}}{(2\pi)^2} \e^{\imag \bs{k}_{K'} \cdot \bs{r}}\int \dd^2 \bs{q} 
  \frac{\imag q_x -  q_y}{q} \e^{\imag \bs{q}\cdot \bs{r}} \notag \\
  &=& -\frac{1}{2} \frac{S_{\rm cell}}{(2\pi)^2} \e^{\imag \bs{k}_{K'} \cdot \bs{r}} 
  \left(\frac{\partial}{\partial r_x} + \imag \frac{\partial}{\partial r_y} \right)
  \int \dd^2 \bs{q}  \frac{1}{q} \e^{\imag \bs{q}\cdot \bs{r}}\notag \\
  &=& 
  - \frac{S_{\rm cell}}{4\pi} 
  \e^{\imag \bs{k}_{K'} \cdot \bs{r}} \frac{r_x + \imag r_y}{r^3}. 
  \label{fromK'}
\end{eqnarray}

\begin{figure}
  \begin{center}
    \includegraphics[width=.8\columnwidth]{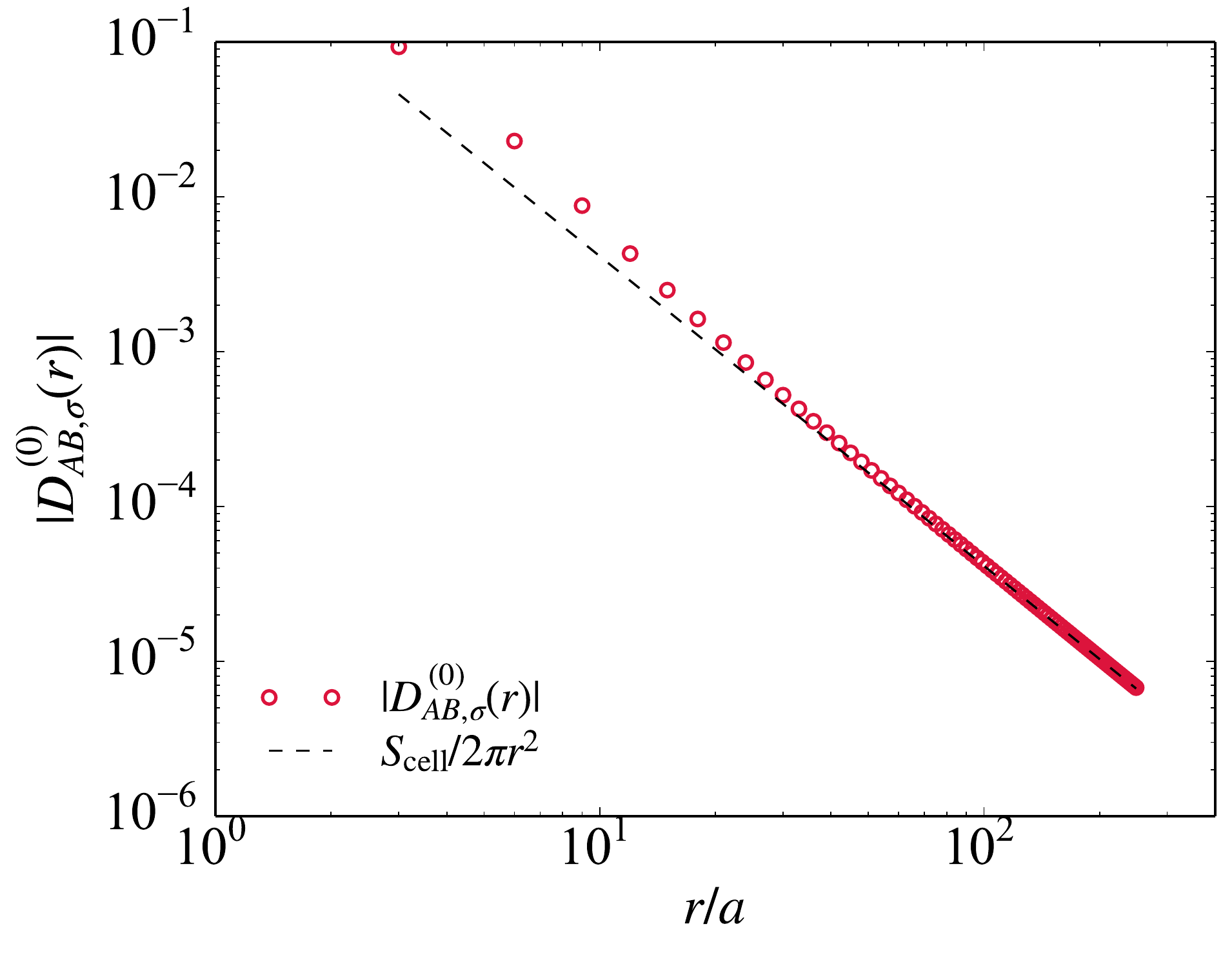}
    \caption{
      \label{fig.D0AB}    
      Log-log plot of 
      $\left |D^{(0)}_{AB,\s}(\bs{r})\right |$ with $\bs{r}=n(\bs{\tau}_1+\bs{\tau}_2) = (3na,0)$   
      calculated directly using Eq.~(\ref{D0sum0})  
      on an $L=1000$ cluster up to $|\bs{r}|/a\leqslant 249$ (red circles). 
      The asymptotic algebraic decay of Eq.~(\ref{D0large_r}) is also shown by dashed line.  
    }
  \end{center}
\end{figure}

The asymptotic form of $D^{(0)}_{AB,\s}(\bs{r})$ for $r/a\gg 1$ is given by 
the sum of (\ref{fromK}) and (\ref{fromK'}), i.e., 
\begin{equation}
  D_{AB,\s}^{(0)}(\bs{r})\simeq -\frac{S_{\rm cell}}{4\pi} 
  \left(
   \e^{\imag \bs{k}_{K}  \cdot \bs{r}} \frac{r_x - \imag r_y}{r^3}
  +\e^{\imag \bs{k}_{K'} \cdot \bs{r}} \frac{r_x + \imag r_y}{r^3} 
  \right).
  \label{fromKandK'}
\end{equation}
Since the contributions from the $K$ and $K'$ points interfere with each other, 
$\bs{r}$ dependence of $D_{AB,\s}^{(0)}(\bs{r})$ is in general complicated. 
Nevertheless, among several directions of $\bs{r}$, 
one can find that $\bs{r}$ in the $x$ direction, i.e., 
$\bs{r}=n(\bs{\tau}_1+\bs{\tau}_2) = (3na,0)$ with $n$ integer, 
gives a simple asymptotic form 
\begin{equation}
  D^{(0)}_{AB,\s}\left(n(\bs{\tau}_1+\bs{\tau}_2)\right) 
  \simeq -\frac{S_{\rm cell}}{2\pi} \frac{1}{r^2} 
  \label{D0large_r}
\end{equation}
for $r/a\gg 1$.
Figure~\ref{fig.D0AB} shows $D^{(0)}_{AB,\s}\left(n(\bs{\tau}_1 + \bs{\tau}_2)\right)$ 
calculated directly on an $L=1000$ cluster using Eq.~(\ref{D0sum0}), which is compared  
with its asymptotic form in Eq.~(\ref{D0large_r}). 
The agreement of the two results for $r/a \gg 1$ verifies 
the algebraic decay of $D^{(0)}_{AB,\s}(\bs{r})$, including the coefficient $S_{\rm cell}/2\pi$.

In the case of an interacting system, it is apparent from Eq.~(\ref{DABFL}) that  
the asymptotic form of $D_{AB,\s}(\bs{r})$ for $r/a\gg1$ 
under the assumption of Eq.~(\ref{Gint}) is given as 
\begin{equation}
  D_{AB,\s}\left(n(\bs{\tau}_1+\bs{\tau}_2)\right) 
  \simeq -Z \frac{S_{\rm cell}}{2\pi} \frac{1}{r^2}. 
  \label{Dlarge_r}
\end{equation}
Therefore, in principle, the quasiparticle weight $Z$ 
can be estimated from the asymptotic behavior of 
the equal-time single-particle Green's function itself, 
without referring to the noninteracting Green's function.

As shown in Fig.~\ref{fig.D0AB}, $D_{AB,\s}(\bs{r})$ of the noninteracting system 
approaches its asymptotic form 
only at a very long distance in a large cluster. 
This might also be the case for the interacting systems. Therefore, 
the direct observation of the asymptotic behavior of $D_{AB,\s}(\bs{r})$ is difficult 
within the cluster sizes affordable at present within the AFQMC method.  
Nevertheless, with an appropriate finite-size-scaling analysis,
we can obtain useful and reliable predictions on the asymptotic behavior,
within the available cluster studied by AFQMC. 
Indeed, we have found that the quasiparticle weight can be estimated more accurately
from the finite-size scaling of the ratio of $D_{AB,\s}(\bs{r})$ between
the interacting and noninteracting systems as in Eq.~(\ref{ratioL}), 
instead of directly fitting the asymptotic behavior of $D_{AB,\s}(\bs{r})$. 
On the other hand, the exponent characterizing the asymptotic behavior of
$D_{AB,\s}(\bs{r})$ in the semimetallic phase can be estimated with 
reasonable accuracy, also for the noninteracting system,
in the way shown in Figs.~\ref{fig.slope} and ~\ref{fig.slope2}.

\section{$D_{AB,\s}(\bs{r})$ in the insulating phase}\label{appB}

In this Appendix, we show that $D_{AB,\s}(\bs{r})$ decays exponentially in $r$
for $r/a\gg 1$ in the insulating phase. 
The derivation is essentially the same as that in Appendix~\ref{appA}.
The main difference due to the finite single-particle excitation gap is that  
the integral over $\bs{q}$ (the momentum measured from the Dirac point),
which yields a massless (Coulomb-potential-like) form for the semimetallic phase as in Eq.~(\ref{Coulomb}), 
now yields a massive (Yukawa-potential-like) form for the insulating phase as in Eq.~(\ref{Yukawa})

To examine the asymptotic form of $D_{AB,\s}(\bs{r})$ 
in the insulating phase, we model the single-particle Green's function $\bs{G}_{\s}(\bs{k},z)$ 
with the same analytical form of 
an antiferromagnetically ordered state, i.e.,  
\begin{eqnarray}
  \bs{G}_{\s}(\bs{k},z) 
  &\approx& 
  \frac{1}{z^2-|\tilde{h}_{\bs{k}}|^2-\Delta^2}
  \left[
    \begin{array}{cc}
      z-(-1)^{\s}\Delta & \tilde{h}_{\bs{k}}\\
      \tilde{h}_{\bs{k}}^* & z+(-1)^{\s}\Delta 
    \end{array}
    \right], 
  \label{Ginsul}
\end{eqnarray}
where $(-1)^{\s} \equiv +1\, (-1)$ for $\s=\up (\dn)$ and $\Delta$ 
is the gap function corresponding to the staggered magnetization 
that breaks the chiral symmetry~\cite{Semenoff2012,Hatsugai2013}. 
Here, we assume that the magnetization is along the $z$ spin-quantization axis with 
real $\Delta\,( >0)$, for simplicity. 
The energy dispersion is obtained by solving $\det\bs{G}_{\s}^{-1}(\bs{k},z)=0$ 
with respect to the frequency $z$, 
i.e., $\pm\sqrt{|\tilde{h}_{\bs{k}}|^2 + \Delta^2}$, and thus it is massive. 
In particular, the single-particle excitation gap at the $K$ and $K'$ points is $2\Delta$.

Inserting the model single-particle Green's function 
into Eq.~(\ref{DAB1}) and taking the zero temperature limit, 
we can obtain the equal-time single-particle Green's function for the insulating phase, i.e.,   
\begin{eqnarray}
  D_{AB,\s}(\bs{r}) 
  &=& -\frac{1}{2} \kint \frac{\tilde{h}_{\bs{k}}}{\sqrt{|\tilde{h}_{\bs{k}}|^2+\Delta^2}} \e^{\imag \bs{k}\cdot \bs{r}}.
\end{eqnarray}
By expanding $\tilde{h}_{\bs{k}}$ around the $K$ point as in Eq.~(\ref{linearK}), 
we find that the contribution to $D_{AB,\s}(\bs{r})$ 
from the momenta around the $K$ point is given as  
\begin{eqnarray}
  &&-\frac{1}{2} \frac{S_{\rm cell}}{(2\pi)^2} \e^{\imag \bs{k}_{K} \cdot \bs{r}}\int \dd^2 \bs{q} 
  \frac{\imag q_x + q_y}{\sqrt{q^2+(\Delta/v_{\rm F})^2}} \e^{\imag \bs{q}\cdot \bs{r}} \notag \\
&=&-\frac{1}{2} \frac{S_{\rm cell}}{(2\pi)^2} \e^{\imag \bs{k}_{K} \cdot \bs{r}} 
  \left(\frac{\partial}{\partial r_x} - \imag \frac{\partial}{\partial r_y} \right)
  \int \dd^2 \bs{q}  \frac{1}{\sqrt{q^2+(\Delta/v_{\rm F})^2}} \e^{\imag \bs{q}\cdot \bs{r}}\notag \\
&=&- \frac{S_{\rm cell}}{4\pi} 
  \e^{\imag \bs{k}_K \cdot \bs{r}} 
  \frac{r_x - \imag r_y}{r^3}
  \left(1-\frac{r\Delta}{v_{\rm F}}\right)
  \e^{-r\Delta/v_{\rm F}},  
  \label{fromKgap}
\end{eqnarray}
where, with the same argument for Eq.~(\ref{Bessel}), the integral over $\bs{q}$ is performed, as in 
the Fourier transform (or the Hankel transform) of the Yukawa potential in two dimensions, i.e.,  
\begin{equation}
  \frac{1}{2\pi}\int \dd^2 \bs{q} \frac{1}{\sqrt{q^2 + (1/\xi)^2}}\e^{\imag \bs{q}\cdot\bs{r}} = \frac{\e^{-r/\xi}}{r} 
  \label{Yukawa}
\end{equation}
with
\begin{equation}
  \xi =\frac{v_{\rm F}}{\Delta}. 
  \label{mfp}
\end{equation}
The propagation of a hole is thus short ranged 
in the insulating phase due to the finite single-particle excitation gap $\Delta$.

With the propagation range $\xi$ of a hole in the insulating phase, 
Eq.~(\ref{fromKgap}) can be written as 
\begin{eqnarray}
- \frac{S_{\rm cell}}{4\pi} 
  \e^{\imag \bs{k}_K \cdot \bs{r}} 
  \frac{r_x - \imag r_y}{r^3}
  \left(1-\frac{r}{\xi}\right)
  \e^{-r/\xi}. 
  \label{fromKgap2}
\end{eqnarray}
Similarly, the contribution to $D_{AB,\s}(\bs{r})$ 
from the momentum around the $K'$ point is evaluated as  
\begin{eqnarray}
- \frac{S_{\rm cell}}{4\pi} 
  \e^{\imag \bs{k}_{K'} \cdot \bs{r}} 
  \frac{r_x + \imag r_y}{r^3}
  \left(1-\frac{r}{\xi}\right)
  \e^{-r/\xi}. 
  \label{fromK'gap}
\end{eqnarray}
Adding (\ref{fromKgap2}) and (\ref{fromK'gap}) yields the asymptotic form 
\begin{equation}
  D_{AB,\s}(\bs{r})\simeq -\frac{S_{\rm cell}}{4\pi} 
  \left(
   \e^{\imag \bs{k}_{K}  \cdot \bs{r}} \frac{r_x - \imag r_y}{r^3}
  +\e^{\imag \bs{k}_{K'} \cdot \bs{r}} \frac{r_x + \imag r_y}{r^3} 
  \right)
  \left(1-\frac{r}{\xi}\right)
  \e^{-r/\xi}. 
  \label{fromKandK'gap}
\end{equation}
In the limit of $\xi \to \infty$, i.e., $\Delta \to 0$, 
Eq.~(\ref{fromKandK'gap}) reduces to 
the noninteracting limit in Eq.~(\ref{fromKandK'}). 
In conclusion, the equal-time single-particle Green's function $D_{AB,\s}(\bs{r})$
decays exponentially in $r$ in the single-particle-gapful system 
with a characteristic length scale $\xi$ given in Eq.~(\ref{mfp}).

\bibliography{bibQP}

\end{document}